\documentclass[preprint,aps, preprintnumbers, nofootinbib]{revtex4}

\newcommand{\gsim}[2]{
\setlength{\unitlength}{12pt}
\begin{picture}(1.4,1.)
\put(.7,-0.3){\makebox(0.0,1.)[t]{$>$}}
\put(.7,-0.3){\makebox(0.0,1.)[b]{$\sim$}}
\end{picture}#2}
\begin{document}

\title{Mass-Varying Neutrinos from a Variable Cosmological Constant}


\author{R. Horvat\footnote{horvat@lei3.irb.hr}}
\affiliation{\footnotesize Rudjer Bo\v{s}kovi\'{c} Institute,
         P.O.B. 180, 10002 Zagreb, Croatia}


\begin{abstract}

We consider, in a completely model-independent way, 
the transfer of energy between the components of the dark energy
sector consisting of the cosmological constant (CC) and that of relic
neutrinos. We show that such a cosmological 
setup may promote neutrinos to  mass-varying
particles, thus resembling a recently proposed scenario of Fardon, Nelson, and
Weiner (FNW), but now without  introducing  any acceleronlike scalar
fields. Although a formal similarity of the FNW scenario with the variable
CC one can  be easily  established, one nevertheless 
finds different laws for neutrino 
mass variation in each scenario. We show that as long as the neutrino number 
density dilutes
canonically,  only a very slow variation of the neutrino
mass is possible. For neutrino masses to vary significantly (as in the FNW
scenario), a considerable deviation from the canonical dilution of the  
neutrino number density is also needed. We note that the present `coincidence'
between the dark energy density and the neutrino energy density can be obtained 
in our scenario even for static neutrino masses.
  
\end{abstract}

\newpage

\maketitle

It is today firmly established experimentally 
that neutrinos have nonzero masses and
nontrivial mixings, thus pointing to the existence of physics beyond the
standard model. In addition, relic neutrinos in the universe, being the
second most abundant particles in the universe, also play an important role
in measuring neutrino mass with cosmological data, in setting bounds
on nonstandard neutrino properties, and also in solving some of the
cosmological problems.  

Another great discovery in recent years has been an observation of the
accelerating rate of expansion of the universe, usually attributed to some
mysterious dark energy sector. The existence of the dark energy sector,
however, constitutes an immediate problem in the form of  today's 
coincidences \cite{1} 
towards the rest of the components in the universe: dark matter, ordinary
matter, radiation,  and neutrinos. From the known behavior of the first three
components since big bang nucleosynthesis (BBN), one finds that any
reasonable tracking of these components by dark energy  
always goes at the  expense
of the late time transition of its equation of state, thus creating a new
problem called the "why now?" problem. The recent idea proposed by Fardon,
Nelson and Weiner (FNW) \cite{2} and developed later by Kaplan, Nelson, and
Weiner \cite{3}, and  
Peccei \cite{4}   
was to circumvent these problems for relic
neutrinos (because the cosmological behavior of their energy density is much
less known), 
by tying together their sector with that of dark energy, in such a
way that dark energy always diluted at the same rate as the neutrino fluid.
This was possible if the mass of the neutrino was promoted to a dynamical
quantity, being a function of the acceleron field (canonically normalized
scalar field similar to quintessence \cite{5}). The main feature of the
scenario \cite{2} is that although the number density of neutrinos dilutes
canonically $(\sim a^{-3})$, the masses of  neutrinos change almost
inversely $(\sim a^{-3\omega })$, thereby promoting their energy density to
an almost undilutable substance. Hence relic neutrinos become tightly
coupled to the original dark energy fluid. If this can be kept for most of
the history of the universe, the near coincidence at present, $\rho_{\Lambda }
\sim \rho_{\nu }$, will cease  to be perceptive as a coincidence at all.

For models with variable mass neutrinos prior to the proposal \cite{2}
see  \cite{6}, and  for those who discussed neutrino mass in
connection with dark energy, see \cite{7}. Application of the FNW proposal 
includes some
studies on leptogenesis \cite{8}, solar neutrinos \cite{9}, and also on the 
cosmo MSW effect \cite{10}.

In the present paper we show how the variable but ``true'' cosmological
constant (CC), with the equation of state (EOS) $\omega_{\Lambda } \equiv
p_{\Lambda }/\rho_{\Lambda }$ being precisely -1, may give rise to 
scale-dependent neutrino masses, in the absence of any acceleronlike scalar 
fields. Throughout the paper we  always  highlight those points in 
which 
our scenario differs from the FNW one.   
Models with the variable CC which 
could successfully mimic quintessence
models and may also shed some light on the coincidence problem (between dark
energy and dark matter), 
have been put
forward recently. Especially relevant are found  those models where the 
CC-variation law was inferred from some underlying physical theory: like
particle physics theory \cite{11, 12}, quantum gravity \cite{13}, or
gravitational holography \cite{14}. A review of phenomenological
CC-variation laws considered before the discovery of  dark energy,
can be found in \cite{15}. The essential features of the cosmological
evolution in variable CC cosmologies were summarized in \cite{16}. 

The continuous transfer of energy between the CC and the gas of 
relic neutrinos (and
{\it vice versa}, depending on the sign of the interaction term) can be
conveniently modeled by the generalized equation of continuity\footnote{For
the sake of simplicity, we
restrict ourselves here to the case of one neutrino family as in
\cite{2}; the generalization to three families of neutrinos is
straightforward.}
\begin{equation}
\dot{\rho }_{\Lambda } + \dot{\rho }_{\nu } + 3H\rho_{\nu } (1+ \omega_{\nu
}) 
= 0 \;,
\end{equation}
where overdots denote time derivatives.
The phase space distribution of relic neutrinos is predicted to be given by
the homogeneous and isotropic Fermi-Dirac distribution of the type
\begin{equation}
f_{\nu }(p)= \frac{1}{\exp\left [{\frac{(p^2 + m_{\nu }^{'2})^{1/2}-\mu_{\nu
}
}{T}}
\right ] + 1} 
\;,
\end{equation}
with $\mu_{\nu } \rightarrow -\mu_{\nu }$ for antineutrinos. Notice that (2)
is not an exact equilibrium distribution because $m_{\nu }^{'} \equiv m_{\nu
}a_{dec}/a $, where $a_{dec}$ denotes the scale factor at the time of
decoupling  
$t_{dec} $. If neutrinos  are relativistic at decoupling $(m_{\nu
} \ll T_{dec} )$, notice that $f_{\nu }$ still retains its relativistic form even
for $T \ll m_{\nu }$, because it is  $m_{\nu }^{'}$ ($m_{\nu }^{'} \ll T$)
and not $m_{\nu }$ that appears in (2). In addition, large neutrino mixing   
revealed in neutrino oscillation experiments may serve to conclude that
chemical potentials for all neutrinos should be small \cite{17}. Hence,
$m_{\nu }^{'}$ and $\mu_{\nu }$ in (2), along with the EOS $\omega_{\nu} 
$ for nonrelativistic neutrinos in (1), can be safely disregarded for all
practical purposes.

We  see from (1) that the CC all the time decays to neutrinos (or
{\it vice versa}), meaning that the energy density of relic neutrinos,
$m_{\nu }n_{\nu }$,  
no longer scales
canonically $\sim a^{-3}$. Here we make a specific {\it anzatz}
 for the energy transfer between the two components coupled through (1),
which leaves the number density of neutrinos to dilute canonically, but
promotes the mass of the neutrino to a running quantity, that is,
\begin{equation}
m_{\nu }(a)=m_{\nu_0 }a^{\alpha }; \;\;\;\;\;\;n_{\nu }(a)=n_{\nu_0 }
a^{-3}\;,
\end{equation}
where $\alpha $ is a constant, the subscript `0' denotes the present-day
values and the present value for the scale factor is set to 1. 
With the {\it anzatz} (3) in (1), one immediately
arrives at
\begin{equation}
n_{\nu } + \frac{\partial \rho_{\Lambda }}{\partial m_{\nu }} = 0\;,
\end{equation}
which is nothing but the assumption of stationarity of $\rho_{dark}$ under
changes in $m_{\nu }$ from the FNW scenario \cite{2}. We thus find that, in
one aspect, the FNW scenario is equivalent to the variable CC
scenario.\footnote{The same conclusion was obtained in \cite{4} but from
considerations which demanded to move away from the nonrelativistic limit.
Here we show that the stationarity condition  for nonrelativistic
neutrinos (4) is a trivial consequence of the continuity relation (1), as
long as the number density of neutrinos scales canonically.} Notice further
that when the interaction phrased by (1) is quenched, {\it i. e.}, when
$\alpha \rightarrow 0$, the {\it anzatz} (3) still keeps neutrino masses
nonzero. It is thus implicitly implied that there exists at least one extra
mechanism (besides that from the CC) for generation of  the neutrino mass. 
This is
not so with the FNW proposal, and it is just this feature which is shown
below  
to be crucial
in determining the law $m_{\nu }(a)$. 

Using the {\it anzatz} (3) it is easy to obtain a solution for $\rho_{\Lambda
}$ in the form
\begin{equation}
\rho_{\Lambda }(a) = \frac{\alpha }{3 - \alpha } \rho_{\nu }(a) + \rho_{\Lambda
}^C\;,
\end{equation}
where $\rho_{\nu }$ (and also $\rho_{\Lambda }$) now scales as 
$ \sim a^{-3 + \alpha }$. 
The integration
constant $\rho_{\Lambda }^C $ is the IR limit of the CC and represents the
true ground state of the vacuum. Regarding Eq. (5), several comments are in
order. If $\alpha > 0$, we are in the realm of decaying CC cosmologies
$(\dot{\rho}_{\Lambda } < 0)$, whereas $\alpha < 0$ means that the transfer
of energy is from neutrinos to the CC $(\dot{\rho}_{\Lambda } > 0)$. 
Since the cosmic matter budget today
consists of no more than $5\%$ of massive neutrinos \cite{18}, one concludes
for the $\alpha > 0$ case that $\rho_{\Lambda }^C $ should always be
nonzero (and positive), unless $\alpha $ is fine-tuned to be very close to 3 
(such large values of $\alpha $ are yet excluded, see below). On the other
hand, for $\alpha < 0$, the first term in (5) is negative, and hence also
the large and positive $\rho_{\Lambda }^C$ is required. These considerations
automatically show that the CC-variation law obtained from gravitational
holography \cite {14} is not able to underpin the present scenario as there
$\rho_{\Lambda }^C$ always vanishes. In addition, we have assumed 
that dark matter and neutrinos
are tightly coupled but also that there is no interaction between dark
energy and baryons (or other components), which means that the Equivalence
Principle is violated. Since the interaction between dark energy and
neutrinos is controlled by the parameter $\alpha $, it is also a measure of
violation of the Equivalence Principle. Next, we adapt the framework
of the effective EOS for the variable CC term (5), in order to obtain the law
$m_{\nu }(a)$.        

The effective EOS for the variable CC whose interaction is phrased by (1)
can be defined similarly as in \cite {19}
\begin{equation}
\omega_{dark}^{eff} = -1 + \frac{1}{3} \frac{d \; ln \; \delta H^2 (z)}
{d \; ln \; (1 +z)}\;,
\end{equation}
where $1+z = a^{-1}$. Here any modification of the standard Hubble parameter
$H$ is encapsulated in the term $\delta H^2 $ (including $\rho_{\Lambda }^C $).
One finally obtains 
\begin{equation}
\omega_{dark}^{eff} = -1 + \frac{(1+z)^{3-\alpha } - (1+z)^3}{\left
(\frac{3}{3-\alpha } \right )(1+z)^{3-\alpha } - (1+z)^3 + \rho_{\Lambda
}^C/\rho_{\nu 0}} \;.
\end{equation}
From (7) one can infer the law $m_{\nu }(a)$ by demanding that
$\omega_{\Lambda }^{eff}$ should not change significantly with $z$. Since
$\rho_{\Lambda }^C/\rho_{\nu 0}$ can be quite large, $\gsim \;15$, one shows
that, for instance, with $d \omega_{dark}^{eff }/d z|_{z=0} \simeq
0.05$,
Eq. (7) can even 
sustain values for $\alpha $ as large as $\gsim \;1$. However, such large
values for $\alpha $ would spoil the tracking behavior at earlier times
when the constant term in the denominator of (7) ceased to be  dominant,
and therefore the only acceptable values are $\alpha \ll 1$. Hence, on
cosmological scales, neutrinos show very slow mass variation. Regarding (7),
one should not be bothered much by the fact that for $\alpha > 0$, the
effective EOS is less than $-1$, since it is well known that arguments
leading to the Big Rip singularity no longer apply  to variable CC models
\cite{19, 16}. 

We have seen that although with the  {\it anzatz} (3) our scenario resembles
the FNW one in that both share the same equation (4) (connecting dark energy
and neutrinos), nonetheless neutrino mass scales quite differently in each 
scenario. To obtain sizable scaling of $m_{\nu }(a)$, one has to enlarge 
the {\it anzatz} (3) to include  the noncanonical scaling of $n_{\nu }$ as
well. We therefore write
\begin{equation}
m_{\nu }(a)=m_{\nu_0 }a^{\alpha }; \;\;\;\;\;\;n_{\nu }(a)=n_{\nu_0 }
a^{-3 +\beta }\;,
\end{equation}
where $\beta $ is a constant. Although, in this case, (4) no longer applies,
one can easily check that the effective EOS is still given by (7), but now
with the replacement $\alpha \rightarrow \alpha + \beta$. The requirement
that $\omega_{\Lambda }^{eff}$ should not vary significantly with $z$ is now
obeyed if $\alpha + \beta \approx 0$. This has strong implications for the
scaling of $m_{\nu }(a)$. One sees that, practically, any law for $m_{\nu
}(a)$ can be underpinned by (7) if the transfer of energy between the two
components is always two-way. For instance, if $\alpha \simeq 3$ (as in the
FNW scenario), one needs $\beta \simeq  -3$. This means that the CC
decays into neutrinos by changing their mass, but simultaneously the neutrino 
component
decays (practically by the same amount) into the CC, thereby changing 
$n_{\nu }$ from its canonical shape. 

Finally, we would like to mention that the variable CC model \cite{12} is
completely able to underpin the present scenario. It is a decaying CC model
with $\alpha + \beta > 0$. The model is based on the renormalization-group
(RG) evolution for $\rho_{\Lambda }$, and on the choice for the RG scale $\mu =
H$. It can be shown that a canonical value for $\alpha + \beta $ is $(4\pi
)^{-1} \ll 1$. In addition, the CC-variation  law, 
$d\rho_{\Lambda }/dz \propto
dH^2/dz$, is a derivative one, thus having a natural appearance of a
nonzero $\rho_{\Lambda }^C$.

We would like to conclude with a few additional comments. It is interesting
to note that even for static neutrino masses $\alpha = 0$, the `coincidence'
$\rho_{\Lambda } \sim \rho_{\nu }$
is still maintained if $\beta \neq 0$ [simply replace $\alpha \rightarrow \beta
$ in (5)]. In addition, the energy density of neutrinos scale almost
canonically in the present scenario ($\alpha + \beta \approx 0$), in strong
contrast with the FNW scenario where the neutrino gas is promoted to an
almost undilutable fluid. This is related to a nice feature that $\rho_{\Lambda
}$, as a solution of (1), always tracks $\rho_{\nu }$, even for small $\alpha
+ \beta $ [cf. (5) with the replacement $\alpha \rightarrow 
\alpha + \beta $]. We are
aware that the assumptions (3) and (8) are not the most general ones, but
also feel that they certainly cover a large number of interesting cases. We
have thus demonstrated how variable CC models may naturally explain the
`coincidence' between dark energy and neutrinos, with both 
neutrino mass changing with the expansion of the universe and without such
changing. Both slow
and quick neutrino mass changing can be naturally implemented in our model.

{\bf Acknowledgments. } The author acknowledges the support of the Croatian
Ministry of Science and
Technology under the contract No. 0098011.


\begin{thebibliography}{160}
\bibitem{1} P. J. Steinhardt, in ``Critical Problems in Physics'', edited by
V. L. Fitch and  Dr. R. Marlow (Princeton University Press, Princeton, N.
J., 1997).
\bibitem{2} R. Fardon, A. E. Nelson and N. Weiner, JCAP 0410 (2004) 005
[astro-ph/0309800].
\bibitem{3} D. B. Kaplan, A. E. Nelson and N. Weiner, Phys. Rev. Lett. 93
(2004) 091801 [hep-ph/0401099].
\bibitem{4} R. D. Peccei, Phys. Rev. D71 (2005) 023527 [hep-ph/0411137].
\bibitem{5} R. R. Caldwell, R. Dave and P. J. Steinhardt, Phys. Rev. Lett.
80 (1999) 1582 [astro-ph/9708069]; 
B. Ratra and P. J. E. Peebles, Phys. Rev. D37 (1988) 3406; C.
Wetterich, Nucl. Phys. B302 (1988) 668.
\bibitem{6} M. Kawasaki, H. Murayama and T. Yanagida, Mod. Phys. Lett. A7
(1992) 563; G. J. Stephenson, T. Goldman and B. H. J. McKellar, Int. J. Mod.
Phys. A13 (1998) 2765 [hep-ph/9603392]; Mod. Phys. Lett. A12 (1997) 2391
[hep-ph/9610317]; R. F. Sawyer, Phys. Lett. B448 (1999) 174
[hep-ph/9809348].
\bibitem{7} R. Horvat, Mod. Phys. Lett. A14 (1999) 2245 [hep-ph/9904451];
JHEP 0208 (2002) 031 [hep-ph/0007168]; 
P. Q. Hung, hep-ph/0010126; P. Gu, X. Wang and X.
Zhang, Phys. Rev. D68 (2003) 087301 [hep-ph/0307148]; A. Brookfield, C. van
de Bruck, D. F. Mota and D. Tocchini-Valentini, astro-ph/0503349.
\bibitem{8} X. J. Bi, P. H. Gu, X. Wang and X. Zhang, Phys. Rev. D69 (2004)
113007 [hep-ph/0311022]; P. H. Gu and X. J. Bi, Phys. Rev. D70 (2004)
063511.
\bibitem{9} V. Barger, P. Huber and D. Marfatia, hep-ph/0502196; M. Cirelli
and M. C. Gonzales-Garcia, hep-ph/0503028.
\bibitem{10} P. Q. Hung and H. Pas, astro-ph/0311131.
\bibitem{11} I. L. Shapiro and J. Sola, Phys. Lett. B475 (2000) 236
[hep-ph/9910462]; JHEP 0202 (2002) 006 [hep-th/0012227]; A. Babic, B.
Guberina, R. Horvat, and H. Stefancic, Phys. Rev. D65 (2002) 085002
[hep-ph/0111207]; B. Guberina, R. Horvat, and H. Stefancic, Phys. Rev. D67
(2003) 083001 [hep-ph/0211184]; A. Babic, B. Guberina, R. Horvat, and H.
Stefancic, astro-ph/0407572.
\bibitem{12} I. L. Shapiro, J. Sola, C. Espana-Bonet and P. Ruiz-Lapuente,
Phys. Lett. B574 (2003) 149 [astro-ph/0303306]; C. Espana-Bonet, P.
Ruiz-Lapuente, I. L. Shapiro and J. Sola, JCAP 0402 (2004) 006
[hep-ph/0311171].
\bibitem{13} A. Bonnano and M. Reuter, Phys. Rev. D65 (2002) 043508
[hep-th/0106133].
\bibitem{14} S. H. Hsu, Phys. Lett. B594 (2004) 13 [hep-th/0403052]; R.
Horvat, Phys. Rev. D70 (2004) 087301 [astro-ph/0404204]; B. Guberina, R.
Horvat, and H. Stefancic, JCAP 05 (2005) 001 [astro-ph/0503495].
\bibitem{15} J. M. Overduin, F. I. Cooperstock, Phys. Rev. D58 (1998) 043506
[astro-ph/9805260]
\bibitem{16} P. Wang, X-He. Meng, Class. Quant. Grav. 22 (2005) 283
[astro-ph/0408495].
\bibitem{17} A. D. Dolgov, S. H. Hansen, S. Pastor, S. T. Petcov, G. Raffelt
and D. V. Semikoz, Nucl. Phys. B632 (2002) 363 [hep-ph/0201287].
\bibitem{18} D. N. Spergel et al., Astrophys. J. Suppl. 148 (2003) 148
[astro-ph/0302209]; M. Tegmark, Phys. Rev. D69 (2004) 103501
[astro-ph/0310723]; U. Seljak et al., astro-ph/0407372; S. Cole,
astro-ph/0501174; D. J. Eisenstein, astro-ph/05001171.
\bibitem{19} E. V. Linder and A. Jenkins, MNRAS 346 (2003) 573
[astro-ph/0305286]; E. V. Linder, Phys. Rev. D70 (2004) 023511
[astro-ph/0402503].
\bibitem{20} R. R. Caldwell, M. Kamionkowski and N. N. Weinberg, Phys. Rev.
Lett. 91 (2003) 071301 [astro-ph/0302506]; E. Elizalde, S. Nojri, S. D.
Odintsov and P. Wang, hep-th/0502082. 
\end{thebibliography}
\end{document}